\documentclass[aps,rmp]{revtex4}
\usepackage{amsfonts,amssymb,amsmath,bm}
\usepackage{graphics,epsfig}
\usepackage{verbatim,ulem}
\usepackage{url}
\usepackage{here}
\usepackage{color}
\usepackage[ngerman,french,english]{babel}

\newcommand{\ue}{\mathrm{e}}

\begin{document}
\title{ROBERT H. KRAICHNAN}
\author{Gregory Eyink}
\affiliation{Dep. Appl. Math. \& Statistics. The Johns Hopkins
  University. Baltimore MD 21218, USA}
\author{Uriel Frisch}
\affiliation{UNS,~CNRS,~OCA,~Lab.~Cassiop\'ee,~B.P.~4229,~06304~Nice~Cedex~4,~France}
\date{\today}
\maketitle

\section{Introduction}

\begin{figure}[h]
\includegraphics[scale=0.8]{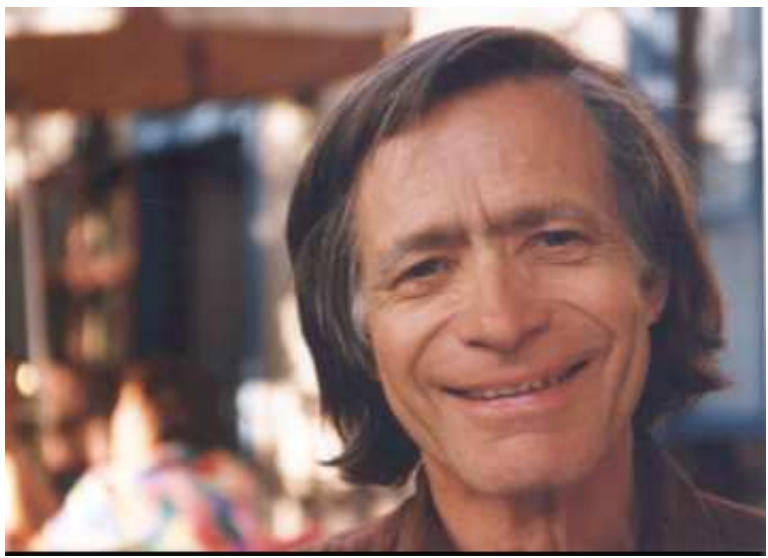}
\caption{Robert H. Kraichnan. Photo by Judy Moore-Kraichnan}
\label{f:rhk}
\end{figure}

Robert Harry Kraichnan {(1928--2008)} was one of the leaders 
in the theory of turbulence for  a span of about forty years (mid-fifties to
mid-nineties). Among his many contributions, he is perhaps best known
for his work on the
inverse energy cascade (i.e. from small to large scales) for forced
two-dimensional turbulence.
This discovery was made in
1967 at a time when two-dimensional flow was becoming increasingly
important for the study of large-scale phenomena in the Earth's 
atmosphere and  oceans. The impact of the discovery was 
amplified by the development of new experimental  and
numerical techniques that allowed full validation of the conjecture.

How did Kraichnan become interested in turbulence? His earliest
scientific interest was in general relativity, which he began to study
on his own at age 13. At age 18 he wrote at MIT a prescient
undergraduate thesis, ``Quantum Theory of the Linear Gravitational
Field''; he received a PhD in physics from MIT in 1949 for his thesis,
``Relativistic Scattering of Pseudoscalar Mesons by Nucleons,''
supervised by Herman Feshbach. His interest in turbulence arose in 1950 while 
assisting Albert Einstein in search for
highly nonlinear, particle-like solutions to unified field equations.
He wrote to his long-time friend and collaborator J.R.~Herring in  the
early nineties how this happened:\\[-0.5ex]
\begin{quote}
I realized I had no real idea of what I was doing and
turned to Navier--Stokes as a nonlinear field problem where experiment
could confront speculation. After initial surprise that
turbulence did not succumb rapidly to field-theoretic attack, I
have been trapped ever since. My overall research theme regarding
turbulence has been to understand what aspects of turbulence
can and cannot be described by statistical mechanics;
that is, what characteristics follow from invariances, symmetries,
and simple statistical measures, and what, in contrast,
can be known only from experiment or detailed solution of the
equations of motion. The principal tool I have used is the construction
of a variety of stochastic dynamical systems that incorporate
certain invariances and other dynamical properties of
actual turbulence but whose statistics are exactly soluble. The
similarities and differences between these model solutions and
real turbulence help illuminate what is essential in turbulence
dynamics. My interests have centered on isotropic Navier--Stokes
and magnetohydrodynamic turbulence in two and three
dimensions.
\end{quote}

When Kraichnan started  working in turbulence, the community was trying to
turn A.N.~Kolmogorov's 1941 (K41) ideas into a quantitative theory devoid
of adjustable parameters. Kraichnan was the only one who truly succeeded in
this endeavor, around the early sixties, by making among other things
extensive use
of his training in theoretical physics and field theory. By that time it had however
become clear that K41 is not the final word, because it misses
intermittency effects that produce anomalous scaling,
that is, scaling laws whose exponents cannot be obtained by dimensional
considerations.  Kraichnan's last major work proposed a fluid-dynamical
framework for studying intermittency that quickly led the younger
generation to identify the mathematical mechanism of
intermittency and also to calculate by perturbation theory  the 
anomalous scaling exponents for Kraichnan's passive scalar
model.\footnote{On K41, cf. Kolmogorov, 1941; Frisch, 1995.}

These were forty difficult years, particularly the first twenty
when K41 seemed to collapse, no \textit{ab initio} theory was emerging
and direct numerical simulation had not yet reached Reynolds numbers
high enough to supplement experimental data on intermittency.
Thus a kind of crossing of the desert took place, with Kraichnan leading
a small flock that found help mostly from geophysicists (at 
the Wood Hole Oceanographic Institute's GFD Program and at the National Center
for Atmospheric Research, Boulder).

This crossing of the desert occurred at a time when Kraichnan 
took
himself ``far from the madding crowd''. After having trod the
traditional path of academia with positions at Columbia University
and Courant Institute, he decided in the very early 60's to become a
self-employed turbulence consultant, solely funded by research
grants.  He moved from New York City to secluded mountains in New
Hampshire, where he lived and worked for almost two decades. 
  During those years he warmly welcomed visitors from all around the
  world and regularly participated in scientific
    meetings, workshops and schools.  Eventually, he moved to New
  Mexico, close to the Los Alamos National Laboratory.  

We shall take the reader on a tour of key contributions to turbulence by Kraichnan,
organized in three main sections, roughly by chronological order:
closure, realizability, the issue of Galilean invariance and MHD turbulence 
(Sec.~\ref{s:closure});
equilibrium statistical mechanics and two-dimensional turbulence
(Sec.~\ref{s:statmech2d}); intermittency and the Kraichnan 
passive-scalar model
(Sec.~\ref{s:intermittency}). In the final section
(Sec.~\ref{s:conclusion}),
we first present two contributions of Kraichnan
which do not fit naturally into the three main scientific sections:
his very first published paper, on scattering of sound by turbulence
(Sec.~\ref{ss:scattering}), and his prediction of the behavior of
convection at extremely high Rayleigh numbers
(Sec.~\ref{ss:convection}). We then turn to Kraichnan's impact on
computational turbulence and present concluding remarks.

The emphasis will be on understanding the flow of ideas and how they
relate to the meandering history of the subject. We shall mostly avoid
technical material. More details may be found in Kraichnan's numerous papers
(over one hundred),  published in journals such as Phys. Rev., Phys. Fluids,
J. Fluid Mech. and J. Atmosph. Sci., and in various reviews and textbooks by other 
authors,\footnote{Leslie, 1973; Monin \& Yaglom, 1975: \S\,19.6; Orszag, 1977; Rose \&
  Sulem, 1978.} but also in 
proceedings papers by Kraichnan, which are often more elementary and 
reader-friendly than the  journal papers.\footnote{Cf. Kraichnan, 1958c, 1972,
1975a.}

A word of warning: in the sequel, standard notions used in turbulence theory, such
as the Navier--Stokes and the magnetohydrodynamics (MHD) equations,
statistical homogeneity and isotropy, energy spectra, etc will
be taken for granted.\footnote{Cf., e.g., Monin \& Yaglom, 1975;
Frisch, 1995.}

\section{Closures: realizability, Galilean invariance and
the random coupling models; MHD turbulence}
\label{s:closure}

After World War II the K41 theory became universally known and was
seen to be mostly consistent with experimental data of that time. 
Attempts were then made to build a quantitative theory of homogeneous and
isotropic turbulence,
compatible with K41 but free of undetermined constants,
such as the (Kolmogorov) constant appearing in the energy spectrum and
able
to predict 
the functional form of the spectrum in the dissipation range. Particularly
noteworthy was S. Chandrasekhar's attempt, based on the quasi-normal approximation
(QNA). Introduced by M.D.~Millionschikov, the QNA assumes that second- and fourth-order moments are related as
in a normal (Gaussian) distribution. 
This is an instance of what is
called \textit{closure}: replacing the infinite hierarchy of moment equations
derivable from the Navier--Stokes equations by a finite number of equations
for suitable statistical quantities such as two-point correlations.
Kraichnan was the first to point out that the simplest closures
that come to mind, such as the QNA can
have \textit{realizability} problems: basic probabilistic inequalities
may be violated. A few years later, using the form of the QNA obtained by 
T.~Tatsumi, Y. Ogura showed indeed that it
can lead to an energy spectrum which is negative for certain
wavenumbers.\footnote{Millionschikov, 1941; Chandrasekhar, 1955;
cf. also  Spiegel, 2010;
Kraichnan, 1957b; Tatsumi, 1957; Ogura, 1963.}

\subsection{The Direct Interaction Approximation and the Galilean
  problem}
\label{ss:dia}

Kraichnan thus set out to obtain better closures, simultaneously free
of adjustable parameters and realizable. He was the first to apply
field-theoretic methods to non-equilibrium statistical fluid
dynamics (Navier--Stokes and MHD): he observed that formal expansions in 
powers of the nonlinearity of the solutions of such equations, followed by averaging 
over random initial conditions and/or driving forces, generates terms
which can be represented by Feynman diagrams.  Eventually,
P.C.~Martin, E.D.~Siggia and H.A.~Rose  would establish a detailed formal connection
between statistical dynamics of classical fields, such as turbulence dynamics, and a 
certain quantum field theory (so-called MSR field theory).\footnote{Kraichnan,
  1958c; 
Martin, Siggia \& Rose, 1973.}

In tackling the closure problem from this point of view, Kraichnan was led
to introduce a new object into turbulence theory, namely the
(averaged infinitesimal) response function, that is, the linear
operator giving the small change in the velocity field at time $t$,
due to a small change in the driving force at times $t'<t$. 
In 1958, using a mixture of field-theoretic considerations and
heuristic but plausible simplifying assumptions, Kraichnan 
proposed the Direct Interaction Approximation
(DIA).\footnote{Kraichnan, 1958a, 1958c, 1959a.} When applied to
homogeneous, isotropic and parity-invariant turbulence, the DIA takes
the
form of two coupled integro-differential equations for the two-point
two-time velocity correlation function and the 
response function. These equations are written  in Sec.~\ref{ss:rcm} in
abstract and concise notation, valid without assuming
any particular symmetry. For homogeneous isotropic turbulence the DIA
equations, which resemble in more complicated form the EDQNM equation
\eqref{preEDQNM} written below,   can
be found in Kraichnan's papers and in the references mentioned in the
Introduction. Such equations
are easily extended to anisotropic and/or helical
(non-parity-invariant) turbulence.

As Kraichnan {observed} himself, the DIA is not
 consistent with K41. In particular the 
inertial-range energy spectrum predicted by DIA is (to leading order)
\begin{equation}
E(k) = C' (\varepsilon v_0)^{1/2} k ^{-3/2},
\label{threehalves}
\end{equation}
instead of
\begin{equation}
E(k) = C \varepsilon ^{2/3} k ^{-5/3},
\label{fivethirds}
\end{equation}
where $\varepsilon$ is the kinetic energy dissipation per mass,
$v_0$ is the r.m.s. turbulent velocity, and $C$ (the Kolmogorov
constant) and $C'$ are dimensionless
constants. The discrepancy has to do with the
way the larger eddies from the energy-containing range of wavenumbers
interact with smaller eddies from the inertial range.  Kraichnan expressed his doubts that  the former merely\\[-1.5ex]
\begin{quote}
convect local regions of the fluid bodily without significantly 
distorting their small-scale (high-$k$) internal
dynamics.
\end{quote}
 This assumption 
would not permit the presence of $v_0$ in the inertial-range expression
of the energy spectrum, except through an intermittency effect
of a kind which was not seriously considered until the work
by the  Russian School in the early sixties. Nevertheless, Kraichnan  
{observed} the following:\\[-1.5ex]
\begin{quote}
As a counter-consideration to Kolmogorov's original argument
in the $x$-space representation, it may be noted that the fine-scale
structure of high Reynolds-number turbulence consists typically
not of compact blobs but of a complicated tangle of extended
vortex filaments and sheets.\footnote{Kraichnan, 1958a: \S\S\,9-10 and 
footnote~\ref{thefootnote}.} 
\end{quote}

For a while Kraichnan {believed} that the DIA may be 
asymptotically exact in the limit $L\to \infty$ for
turbulence endowed with  $L$-periodicity in all the  spatial coordinates.
However, after a few months, Kraichnan {found} 
that the DIA equations are actually the exact consequences of a certain
random coupling model obtained from the hydrodynamical equations (Navier--Stokes or
MHD) by a suitable modification of the nonlinear interactions (see Sec.~\ref{ss:rcm}). The presence
of this model automatically guarantees the realizability of the DIA. The random
coupling model has many properties in common with the original hydrodynamical
equations, such as symmetries,  energy conservation, etc. But Kraichnan
{noted} that, when various correlation functions are expanded in terms
of Feynman diagrams, the random coupling model retains only a subclass of 
all diagrams; in particular it misses the so-called vertex corrections
which contribute (in field-theoretic language) to the renormalization 
of the nonlinear interaction.\footnote{Kraichnan, 1958b, 1958c.}

Six years later he {discovered} a more serious defect:
the random coupling model and the DIA fail invariance
under \textit{random Galilean transformations}.\footnote{Kraichnan, 1964b;
Note that the random coupling model can be modified to preserve
a Galilei symmetry group of dimension 3, for a model with $N$ 3-dimensional velocity 
fields. See Kraichnan, 1964a: Appendix. However, this is insufficient to guarantee 
random Galilean invariance for the DIA equations.} Ordinary Galilean invariance
--- for the Navier--Stokes equations --- is the observation that if 
$({\bm u}({\bm x},t),\,p({\bm x},t))$ are the velocity and pressure fields
which solve the Navier--Stokes equations in the absence of boundaries
or with periodic boundary conditions, then 
$({\bm u}({\bm x} -{\bm V}t,t)+ {\bm V},\,p({\bm x} -{\bm V}t,t))$ are also  solutions for
an arbitrary choice of the velocity ${\bm V}$. Random Galilean invariance has 
the velocity ${\bm V}$ chosen randomly with an isotropic distribution and
independent of the turbulent velocity-pressure fields. Isotropy ensures
that the mean velocity remains zero. In the DIA, when a random Galilean
transformation is performed, the mean square velocity $v_0^2$ increases
and thus the spectrum given by \eqref{threehalves} changes. On the one
hand this is clearly
inconsistent with the Galilean invariance of the Navier--Stokes equations. On the 
other hand, an influence of the energy-carrying eddies on the inertial range
different from what K41 predicts {(e.g. intermittency effects)} cannot be ruled out. 

But first, Kraichnan {realized} that the DIA had to be modified to restore random
Galilean invariance and, if possible, without
losing its other nice features. To explain how this can be done,
we have to become slightly more technical. Kraichnan derives
the following equation from the DIA for the evolution of
the energy spectrum when the turbulence is homogeneous, isotropic and
parity-invariant:
\begin{eqnarray}
&&\left(\partial_t +2\nu k^2\right)
E(k,t)  
= 
\int\int_{\triangle_k}dpdq\,\theta_{kpq}\,\times \nonumber \\[1.6ex]
&& b(k,p,q)\frac{k}{pq}E(q,t)\left[k^2E(p,t)-p^2E(k,t)\right]+F(k)\;,
\label{preEDQNM}
\\[1.6ex]
&&\theta_{kpq}\equiv\frac{(\pi/2)^{1/2}}
{v_0(k^2+p^2+q^2)^{1/2}}\;, \quad b(k,p,q)=\frac{p}{k}(xy+z^3)\;.
\label{defthetaandb}
\end{eqnarray}
Here $E(k,t)$ is the energy spectrum, $F(k)$ the energy input from
random driving forces, $\triangle_k$ defines the set
of $p\ge 0$ and $q\ge 0$ such that $k$, $p$, $q$ can form 
a triangle, 
$x$, $y$, $z$ are the cosines of the  angles opposite to sides $k$,
$p$ and $q$ of this triangle. The factor $\theta_{kpq}$ can be
interpreted as a triad relaxation time. The particular form given in
\eqref{defthetaandb} is obtained from the DIA by an asymptotic
expansion valid only when the wavenumber $k$ is in the inertial range.
This form implies that relaxation has a time scale comparable to the
convection time $1/(kv_0)$ of an eddy of size $1/k$ at the r.m.s.
velocity $v_0$. As Kraichnan notes, if $v_0$ is replaced
``by, say, $[kE(k)]^{1/2}$, which may be considered the r.m.s. velocity
  associated with wavenumbers the order of $k$ only'', then K41 is
  recovered. Actually this choice, which also restores random
Galilean invariance and is easily shown to be realizable, became later known as
the Eddy Damped Quasi-Normal Markovian (EDQNM)
approximation.  Note that the EDQNM requires the use of an adjustable
dimensionless constant in front, say, of $[kE(k)]^{1/2}$, which can
  then be tuned to give a Kolmogorov constant matching experiments
 and/or simulations. 
Kraichnan clearly preferred having a systematic
theory free of such constants and then seeing how well it can
reproduce experimental data, not just constants, but also the complete 
functional forms of spectra.  He  eventually developed a
version
 of the EDQNM, called the Test Field Model in which the triad
 relaxation time is determined in a systematic way through the
 introduction
of a passively advected (compressible) test velocity field and a
procedure for eliminating unwanted sweeping
effects.\footnote{Kraichnan, 1958c; on EDQNM, cf. Orszag, 1966, 
1977 and see also Rose \& Sulem, 1978; Lesieur, 2008; on the 
Test Field Model, cf. Kraichnan, 1971a, 1971b.}

To overcome the difficulty with random Galilean invariance in a more
systematic way, however, he first developed the Lagrangian History Direct
Interaction Approximation (LHDIA) and an abridged version thereof (ALHDIA) 
in which the correlation functions have fewer time arguments. LHDIA and 
ALHDIA make
use of a generalized velocity ${\bm u}({\bm x},\,t|t')$ which has both
Eulerian and Lagrangian characteristics.  It is defined as the
velocity measured at time $t'$ in that fluid element which passes
through ${\bm x}$ at time $t$, and it satisfies an extended form of
the Navier--Stokes equation in the two time variables. When the DIA
closure is written for this extended system there appear integrals
over the past history of the flow, which can be altered when working
with the generalized field in such a way as to recover random Galilean
invariance.  Detailed numerical studies of the ALHDIA gave a full
functional form of the spectra in very good agreement with
experimental data.  In particular the Kolmogorov constant was $C=1.77$
while the best current value is 1.58, when ignoring intermittency corrections. Other detailed predictions were
made for various Lagrangian quantities involving correlations of
velocity increments, pressure-gradients, particle accelerations and
also for 2-particle dispersion. Many of these predictions are still to
be subjected to experimental test.\footnote{Kraichnan, 1964b, 1965a,
  1966a; on experimental data, cf. Grant, Stewart \& Moilliet, 1962;
  on the Kolmogorov constant, cf. Donzis \& Sreenivasan,
  2010.  Ott \& Mann, 2000, corrected an error in Kraichnan's
  LHDIA theory of turbulent two-particle dispersion (Kraichnan, 1966b) 
  and found from experiment that its prediction for the Richardson-Obukhov 
  constant was too large by a factor of 10.}

In the course of studying the DIA and all its ``children'', Kraichnan
had to develop quite sophisticated analytic and numerical tools. For
example he showed that the spectrum falls off exponentially
in the far dissipation range and obtained the algebraic prefactor in
front of the exponential.
Such tools, are transposable almost immediately to 
the study of other closures, for example of the kinetic theory
of resonant nonlinear wave interactions.\footnote{Zakharov, L'vov \&
Falkovich, 1992; Nazarenko, 2010
.} They are also part of the Kraichnan legacy.

Many of Kraichnan's tools from statistical mechanics and field theory
were unfamiliar to fluids scientists trained in the traditions of
theoretical mechanics and applied mathematics, and some objects
central to the DIA ---particularly the mean response function---did
not appear in previous exact moment hierarchies. The DIA was thus
received with some scepticism by the scientific community, in
particular in Cambridge (UK), but Kraichnan was given the opportunity
to present his theory in the recently introduced Journal of Fluid
Mechanics.  Kraichnan's theory was discussed by Proudman at the famous
Marseille 1961 conference. With hindsight,  we see 
that he failed to notice the realizability of the DIA and lumped it 
together with the quasi-normal approximation (QNA); actually, it would
take two more years until Ogura discovered the aforementioned problems
with QNA.  Much more interesting is that, at the end of his paper,
Proudman (correctly) expressed doubts that DIA could cope with what was later
termed dissipation-range intermittency, which would be first explained
by Kraichnan six years later
(cf. Sec.~\ref{ss:dri}).\footnote{Kraichnan, 1959a; Proudman, 1961;
  Ogura, 1963.}
Eventually, Kraichnan's efforts to develop an analytical
theory, compatible with K41 and able to give detailed quantitative
predictions, was remarkably successful and influential.  Of course
around the same time, at the Marseille conference, it became
increasingly clear that K41 was not the final word for fully developed
turbulence.  Kraichnan's contributions to post-K41 theories will be
described in Sec.~\ref{s:intermittency}.

\subsection{The random coupling models and the $1/N$ expansion}
\label{ss:rcm}

Starting from the Navier--Stokes (NS) equations Kraichnan derived the DIA,
using heuristic assumptions, that even if plausible, turned out to be
partially false. Quite rapidly after this, Kraichnan showed that the DIA
equations are actually the \textit{exact} consequence in a certain limit of a
stochastic model, called the random coupling model (RCM), which is obtained by
a suitable modification of the Navier--Stokes equations. This is important not
only because it ensures the realizability of the DIA but because it makes
contact with---and largely anticipates---techniques still being developed in
field theory and statistical mechanics. We shall try here to give a
not-too-technical presentation of the RCM.\footnote{Kraichnan, 1958c, 1961.}

Random coupling models can be written for a large class of
quadratically nonlinear partial differential equations, which
encompass the Burgers equation, the Navier--Stokes equations
and the MHD equations. All these can be written in the following
general form, where it is assumed that the solution is zero for $t<0$:
\begin{equation}  
  \partial_t u(t) + B(u(t),u(t)) = Lu(t) +f(t) +u(0)\delta(t).
\label{genquadr}
\end{equation}
Here $u(t)$ (scalar or vector) collects all the dependent variables,
$B(.,.)$ is a quadratic form (comprising the inertial and pressure
terms for the case of Navier--Stokes), $L$ is a time-independent
linear operator 
(e.g., the viscous dissipation operator) and $f(t)$ is a prescribed zero-mean random
force, taken Gaussian for convenience; the prescribed zero-mean initial
condition  $u(0)$ in front of the Dirac distribution $\delta(t)$ is also taken random Gaussian.
Obviously, \eqref{genquadr} can be rewritten  {as an equivalent integral 
equation}
\begin{equation}  
u(t) +\int_0^tdt'\,\ue ^{L(t-t')}B(u(t'),u(t'))=\int_0^tdt'\,\ue
^{L(t-t')}f(t') +u(0).
\label{integralform}
\end{equation}

In somewhat more abstract form this may be written as
\begin{equation}  
u+{\cal B}(u,u) = {\cal F}.
\label{absquadr}
\end{equation}

Let us now describe the random coupling model in the form
used by Herring and Kraichnan.
Imagine that $N$ independent replicas of \eqref{absquadr} are written
with $N$ independent fields, labeled $u_\alpha$  ($\alpha = 1,\ldots,N$) and that
these are then coupled as follows:
\begin{equation}  
 u_\alpha+\frac{1}{N}\sum_{\beta,\gamma}
  \phi_{\alpha\beta\gamma}
 {\cal B}(u_\beta,u_\gamma) = {\cal F}_\alpha.\qquad \alpha
  =1,\ldots, N.
\label{defrcm}
\end{equation}
Here, the ${\cal F}_\alpha$ are $N$ independent identically distributed (iid)
replicas of the force ${\cal F}$ in \eqref{absquadr}; the ``random
coupling coefficients'' 
$\phi_{\alpha\beta\gamma}$ are a set of Gaussian random variables of 
zero mean and unit variance. These $N^3$ coefficients are taken all
independent, except for the requirement that
$\phi_{\alpha\beta\gamma}$  be invariant under all permutations
of $(\alpha,\,\beta,\,\gamma)$, a condition crucial for 
conservation of energy and other quadratic invariants. 
When the
number of replicas tends to infinity, closed equations emerge
for the two  following objects: the correlation function
$ U\equiv \langle u_\alpha \otimes u_\alpha \rangle$ and the infinitesimal response 
function  $ G \equiv \langle \delta u_\alpha/ \delta {\cal
  F}_\alpha\rangle$. Note that off-diagonal terms with $\alpha\neq \beta$
 tend to zero as $N\rightarrow\infty.$
The angular brackets denote ensemble averages and 
$\otimes$ indicates a tensor product: if we were dealing with 
a velocity field $u_i({\bm x},t)$, then $u\otimes u$  would stand for 
$u_i({\bm x},t)u_j({\bm x'},t')$. It is also convenient to introduce
auxiliary independent zero-mean Gaussian random fields ${\hat v}$
and  $v$ having the same correlation function $U$
as any of the $u_\alpha$'s.
With such abstract notation the DIA equations take the compact
form:
\begin{eqnarray}
&&U= \langle (G {\cal F}) \otimes (G {\cal F})\rangle +2G \langle {\cal
  B}({\hat v},{ v})\otimes G {\cal
  B}({\hat v},{ v}) \rangle \\
&&G -4 \langle {\cal B} ({\hat v}, G{\cal B}({\hat v},G))\rangle =I,
\label{abstractdia}
\end{eqnarray}
where $I$ is the identity operator.\footnote{On the random coupling
model, cf. Kraichnan, 1958c, 1961, Herring \& Kraichnan, 1972; 
on the compact form of the DIA,
cf. Lesieur, Frisch \& Brissaud, 1971.}

The random coupling model can also be modified to give equations
other than DIA.  For example the phases $\phi_{\alpha\beta\gamma}$ can
be made 
functions of the time 
with a white noise dependence: $\langle \phi_{\alpha\beta\gamma} (t)
\phi_{\alpha\beta\gamma}(t')\rangle =\tau_0 \delta(t-t')$; this is
the so-called Markovian Random Coupling Model (MRCM)  which leads to  an equation similar to
\eqref{preEDQNM} but with $\theta_{kpq} \equiv \tau_0/4$. The EDQNM equation can also be obtained in
similar manner by
taking the random phase dependent on triads of nonlinearly interacting
wavevectors and delta-correlated in time with a  coefficient
proportional
to the triad relaxation time $\theta_{kpq}$ chosen in agreement with
K41, as explained in Sec.~\ref{ss:dia}. Unfortunately, so far
no random coupling model has been found for the Lagrangian
variants of DIA.\footnote{On variants of the random coupling model,
  cf. Kraichnan, 1958c, 1961; on the MRCM, cf. Frisch, Lesieur
\& Brissaud, 1974.}

Kraichnan's work on the RCMs has many other
interesting connections in mathematics and physics. For example, the
simplest dynamics considered in Kraichnan's 1961 paper is the
classical harmonic oscillator with a random frequency.  The
$N$-replica model in this case reduces to the study of a class of
random $N\times N$ matrices (Toeplitz matrices with coefficients whose
phases are randomly altered, consistent with Hermiticity). Following a
suggestion of Kraichnan, it was shown by U. Frisch and 
R. Bourret that the DIA equation for the harmonic
oscillator can also be obtained from an RCM  employing random Wigner
matrices selected out of the Gaussian Unitary Ensemble. This
approach works for any linear dynamics with a random linear operator,
such as the problems of turbulent advection of a passive scalar or the
Schr\"odinger equation with random potential.  

Kraichnan's discovery
of the RCM in 1958 anticipated, in fact, decades-later developments in
quantum field theory employing large-$N$ models with quenched random
parameters and random matrices. $SU(N)$ gauge field theories for
$N\rightarrow\infty $ were shown by G.~'t~Hooft to resum a subset of all
Feynman diagrams, the planar diagrams, and Yu.M.~Makeenko and 
A.A.~Migdal showed that the exact Schwinger--Dyson equations for 
Wilson loops reduce in that limit to a closed set of self-consistent equations 
like DIA . The most direct connection with Kraichnan's work is for random 
matrix models of large-$N$ gauge theory in 0D and quantum gravity in 2D, 
where the self-consistent ``loop equation'' is identical in form to the DIA
equation for the harmonic oscillator with a non-Gaussian
random frequency.\footnote{Kraichnan, 1961: \S~9.}
Kraichnan's RCM was not a rote application of standard field-theoretic techniques 
of the 1950's, but employed advanced ideas that were regarded as ``cutting edge'' 
decades later.\footnote{On large-N gauge theory and quantum gravity,
  G.~'t~Hooft, 1974;
Makeenko \& Migdal, 1979; Migdal, 1983; Di Francesco, Ginsparg \& Zinn-Justin, 
1995.}

The Random Coupling Models may have more than just historical
interest, however.  One of the successes of theoretical physics in the
1970's was the development of $1/N$ expansion methods to provide
analytical tools to tackle nonperturbative problems with no other
small parameter, such as anomalous scaling in critical
phenomena and quark confinement in quantum
chromodynamics. As we shall discuss in more
detail in Sec.~\ref{s:intermittency}, turbulence anomalous scaling due to
inertial-range intermittency is exactly this sort of
problem. Moreover, recent numerical studies of large-$N$ Random
Coupling Models for some simple dynamical systems model of turbulence,
so-called ``shell models'', have found that inertial-range anomalous
dimensions vanish proportional to $1/N$. 
Kraichnan's pioneering work on the RCM together with modern $1/N$
expansion methods might provide a powerful tool to address the difficult
problem of inertial-range turbulence scaling.\footnote{On DIA and
Wigner matrices, cf. Frisch \& Bourret, 1970. 
On 1/N expansion in critical phenomena, cf. Abe, 1973. 
On large-$N$ shell models,
cf. Pierotti, 1997; Pierotti, L'vov, Pomyalov \& Procaccia, 2000.}.

\subsection{MHD turbulence}
\label{ss:mhd}

Magnetohydrodynamics (MHD) was born with Alfv\'en's prediction of 
a new type of waves (now called Alfv\'en waves) in a conducting fluid in the presence of 
a uniform background magnetic field.   As pointed out by von
Neumann in his report on turbulence at the conference ``Problems of
Motion of Gaseous Masses of Cosmic Dimensions,'' 
MHD was considered key in
understanding the origin of various cosmic magnetic fields --- beginning
with that of the Earth --- and the mechanism for accelerating cosmic
rays. Immediately the question arose if MHD turbulence at high
kinetic and magnetic Reynolds numbers can be described using K41. The
question was difficult because (i) there were no experimental data on MHD
turbulence in the late forties and (ii) dimensional analysis was of
limited use, due to the presence of two fields, the velocity field
${\bm u}$ and the magnetic field ${\bm b}$.\footnote{The magnetic field
can be given the same dimension as the velocity field after division
by $\sqrt{4\pi \mu \rho}$ where $\mu$ is the magnetic permeability and
  $\rho$ the density; these are the units used here.} Important early papers on the statistical
theory of turbulence were already dealing with MHD turbulence, even
in their titles. This included Batchelor's paper on the analogy between
vorticity and magnetic fields, Lee's 1952 paper (cf. Sec.~\ref{s:statmech2d})
and Kraichnan's very first paper on the DIA, where he
also handled the MHD case, but without explicitly stating that
it leads to a $k^{-3/2}$ inertial-range spectrum, as in the purely
hydrodynamic case.\footnote{Alfv\'en, 1942; von Neumann, 1949;
  Batchelor, 1950; Lee, 1952;
  Kraichnan, 1958a.}

One year after discovering  that the DIA fails random
Galilean invariance, Kraichnan returned to MHD and found that 
the $k^{-3/2}$ law may actually apply to the three-dimensional MHD case
when there
is a significant random large-scale magnetic field of r.m.s. strength
$b_0$. In a short research note he
stated:
\begin{quote}In the present hydromagnetic case, it still may be argued plausibly
that the action of the energy range on the inertial range is equivalent to
that of spatially uniform fields. But, in contrast to a uniform velocity
field, a uniform magnetic field has a profound effect on energy
transfer.
\end{quote}
Indeed, when the equations of MHD are rewritten in terms of the Els\"asser
variables ${\bm z}^\pm \equiv {\bm v} \pm {\bm b}$, and a uniform background
field of strength $b_0$ is present, it is found (i) that in linear theory
${\bm z}^+$ and ${\bm z}^-$ propagate as Alfv\'en waves in opposite directions with speed $b_0$ along the
background field, (ii) the only nonlinear interactions are between 
${\bm z}^+$ and ${\bm z}^-$ (no self-interactions). It follows that a 
${\bm z}^+$-eddy and a ${\bm z}^-$-eddy of wavenumbers $\sim k$ will
be able to significantly interact only during a coherence time $\sim
1/(kb_0)$. Thus the flux $\varepsilon$ of total (kinetic plus
magnetic) energy 
through the wavenumber $k$ will be proportional to $1/b_0$, where $b_0$ is the
r.m.s. magnetic field fluctuation (stemming from the energy range).
Therefore $\varepsilon$ and $b_0$ must appear in the combination
$\varepsilon b_0$. It then follows by dimensional analysis that both
the kinetic and magnetic energy spectra are of the form
\begin{equation}
E(k) \sim (\varepsilon b_0)^{1/2} k^{-3/2}.
\label{mhdrhk}
\end{equation}
This is the same functional form as the DIA spectrum \eqref{threehalves},
and for a good reason: the large-scale magnetic field plays now the same role
as the large-scale velocity field in the DIA. Of course, in the DIA this is
a spurious role, whereas in MHD Alfv\'en waves make this effect real.
It must be pointed out that Iroshnikov  proposed the same
MHD spectrum \eqref{mhdrhk} slightly before Kraichnan, using a semi-phenomenological
theory of interaction of three Alfv\'en waves and a diffusion approximation in
$k$-space
for the energy spectrum derived by arguments  resembling those  Leith 
used for two-dimensional turbulence. 
Iroshnikov's work did not penetrate much into
the West until approximately 1990. Nowadays the energy spectrum
\eqref{mhdrhk} is called the Iroshnikov--Kraichnan
spectrum.\footnote{Kraichnan, 1965b; Iroshnikov, 1963; Leith 1967.}

Finally we mention that the arguments of Iroshnikov and Kraichnan
can be questioned because local anisotropies are ignored: in a small region
in which the (large-scale) magnetic field is ${\bm B}$, those small-scale
eddies having wavevectors perpendicular to ${\bm B}$, or nearly so, are hardly
affected by Alfv\'en waves. Hence some kind of quasi-two-dimensional MHD
turbulence can emerge. A more systematic theory of weakly interacting
Alfv\'en waves predicts a $k^{-2}$ spectrum for MHD
turbulence. The problem of determining the spectrum of {strong} 
MHD turbulence---even at the level of a Kolmogorov mean-field theory, ignoring 
intermittency---is still quite open at
this time.\footnote{On anisotropies, cf. Kit \& Tsinober, 1971;
R\"{u}diger, 1974;  Shebalin, Matthaeus, \& Montgomery;
1983. On ideas of using resonant wave interactions, cf. Ng \&
Bhattacharjee, 1996; Goldreich \&
Sridhar, 1997. On the systematic theory of resonant wave interactions,
cf. Galtier {\it et al.}, 2000; Nazarenko, Newell \& Galtier, 2001.}

\section{Statistical mechanics and two-dimensional turbulence}
\label{s:statmech2d}

The essentially statistical character of turbulent flow has been
clearly understood long before even the 1941 work of Kolmogorov, by
such scientists as 
Lord Kelvin,
G.I.~Taylor, N.~Wiener, T.~von K\'arm\'an, and
J.M.~Burgers. To this day, probably the most successful statistical
theory of any dynamical problem is the equilibrium statistical
mechanics of Gibbs, Boltzmann and Einstein for classical and quantum
Hamiltonian systems.
It was natural then that the pioneers in turbulence
theory would seek some guidance and inspiration from Gibbsian
statistical mechanics.  Notable is the attempt by Burgers to apply the
maximum entropy principle to turbulent flows in a 
series of papers in 1929--1933. Unfortunately, it was pointed out by Taylor in
1935 that mean energy dissipation (and thus
entropy production) does not vanish in the limit of high Reynolds
numbers for typical turbulent flows.
Turbulence is thus
a fundamentally non-equilibrium problem to
which the Boltzmann--Gibbs formalism is not \textit{directly} applicable.
Nevertheless, a judicious application of equilibrium statistical theory to hydrodynamics 
can still yield crucial hints and ideas on the behavior of turbulent flows and Kraichnan was 
one of the most masterful practitioners of this art. He used Gibbsian statistical predictions 
in his construction of closure theories and, most subtly, to help to divine the behavior of
strongly disequilibrium turbulent cascades. This approach played an especially important 
role in Kraichnan's development of his dual cascade picture of two-dimensional 
turbulence. For this reason, we shall treat these two subjects together. 

\subsection{Equilibrium Statistical Hydrodynamics}
\label{ss:statmech}

Prior to Kraichnan's work, one of the most significant applications of
Gibbsian statistical mechanics to hydrodynamics was in fact to
two-dimensional fluids, by L. Onsager in a 1949 paper titled
``Statistical Hydrodynamics''.  Onsager noted
that ``two-dimensional convection, which merely redistributes
vorticity'' leads to energy conservation at infinite Reynolds
number. Thus, equilibrium statistical mechanics may be legitimately
applied to 2D Euler hydrodynamics, if 
ergodicity is assumed (as usual). Onsager studied in detail the microcanonical
distribution of a ``gas'' of $N$ Kirchhoff point-vortices in an
incompressible, frictionless, 2D fluid. His most striking conclusion
was that, for sufficiently high kinetic energies of the
point-vortices, thermal equilibrium states of \textit{negative} absolute
temperature occur.  The point-vortices condense into a single
large-scale coherent vortex, which Onsager suggested could explain the
``ubiquitous'' appearance of such structures in quasi-2D 
flows.\footnote{Onsager, 1949.}

A conservative system appears in any space dimension if the Euler
 equations or the MHD equations for ideal unforced fluid flow are
 Galerkin truncated, as first noted by Burgers and, later
 independently, by T.D.~Lee and E.~Hopf.  For the case of
 space-periodic flow and in the abstract notation of Sec.~\ref{ss:rcm}
 this amounts to replacing 
the nonlinear term in (\ref{genquadr}) by $P_{\rm G}B(u(t),u(t)),$ 
where $P_{\rm G}$ is the Galerkin projection. This operator is defined in terms
of the spatial Fourier representation of $u(t)$ by setting to zero all Fourier
components  of wavenumbers $k> K_{\rm G}$, the truncation wavenumber. 
The above authors showed then
that, in suitable variables (based on the real and imaginary parts
of the the Fourier amplitudes) 
the dynamics can be written as a set of ordinary real differential
equations
\begin{equation}  
\dot q_\alpha = F(q_1,\ldots,q_N), \qquad \alpha = 1,\ldots, N, 
\label{ode}
\end{equation}
with a Liouville theorem
$\sum_\alpha \partial \dot q_\alpha/\partial q_\alpha =0,$  
which expresses the conservation of volumes in the $N$-dimensional
phase space. The hypothesis of ergodicity 
implies equipartition of energy among all degrees of freedom. In three
dimensions, this led Lee and Hopf to a $k^2$ energy spectrum, with no
divergence of the total energy because of truncation.  Both
recognized that, with viscosity added and absent truncation,
things will be very different and that Kolmogorov's $-5/3$ spectrum is
expected.  In the MHD case, Lee obtained the same spectrum for the
magnetic field and an equipartition between kinetic and magnetic
energy for every Fourier harmonic. He conjectured this extends to
non-equilibrium MHD turbulence, inferring that 
Kolmogorov's $-5/3$ spectrum should also hold in the MHD 
case.\footnote{Hopf, 1952;  Lee, 1952.}

This was roughly the situation when Kraichnan entered the field of
turbulence, and statistical mechanical ideas dominated much of his
early thinking, too.\footnote{Kraichnan had discussions about
statistical mechanics and the foundations of thermodynamics 
in the late fifties with B.~Mandelbrot (Mandelbrot, 
private communication, 2010).}
His first work on the application of equilibrium
statistical mechanics to turbulence was in a hardly cited 1955 paper on
compressible turbulence in which he rediscovered, in a slightly
different context, the 1952 results of Lee.
In particular Kraichnan found
the Liouville theorem and equipartition solutions (here, the
equipartition is between kinetic and potential acoustic energies).
Kraichnan's preoccupation with statistical mechanics continued in his first 
DIA paper.  He discussed there the Gibbs ensembles for the
Galerkin-truncated equations but noted that ``This artificial
equilibrium case does {\it not} describe turbulence at infinite
Reynolds number,'' when a forcing term  and a viscous damping term
are added. 

 There is, however, one very original and definitive result on equilibrium statistical mechanics 
obtained in Kraichnan's first paper on DIA. He established there a Fluctuation-Dissipation 
Theorem (FDT; called by him, correctly, ``Fluctuation-Relaxation'') for conservative nonlinear 
dynamics in thermal equilibrium, generalizing previous results of H.B.~Callen and co-workers.  
Kraichnan's derivation in an Appendix to the 1958 paper was very ingenious. 
He showed that the FDT arises from the stability of the Gibbs 
measure for two replicas of the original dynamics under couplings that preserve 
the Liouville theorem and energy conservation. Recognizing the more general interest
in this result, Kraichnan one year later published this argument separately for any
dynamics that conserves both phase volume and an arbitrary energy 
function $E.$ In the special case of the truncated Euler system 
considered in the 1958 DIA paper, 
Kraichnan's FDT reduces to the statement that the 2-time correlation function $U$ and 
the mean response function $G$ are proportional, with the absolute temperature supplying 
the constant of proportionality. In this case, the DIA equations reduce to a single equation 
for $G,$ which Kraichnan discussed in a separate section on ``Nondissipative Equilibrium'' 
in his paper.  This is the same as the ``mode-coupling theory'' applied to equilibrium 
critical dynamics twelve years later by K. Kawasaki (who was aware of and 
influenced by Kraichnan's earlier work on DIA).\footnote{Kraichnan, 1958a, 1959b, Kawasaki, 1970.} 

Kraichnan returned to the statistical mechanics of 
Galerkin-truncated ideal flow in a 1975 article ``Remarks on Turbulence Theory.''
Although ostensibly  a review article, 
this work---characteristically for Kraichnan---contained many original ideas and results 
not published elsewhere. Kraichnan pointed out that the ``absolute equilibrium'' 
for the inviscid truncated system in 3D had an interesting domain of physical applicability, 
describing the small thermal fluctuations of velocity in a fluid at rest.
As an application of the equilibrium DIA equations, he showed that
\begin{quote}
\ldots the truncated Euler system in thermal equilibrium exhibits a dynamical
damping of low-wavenumber disturbances just like the viscous damping of the
Navier--Stokes system at zero temperature. If $k_{\max}$ is taken as some kind of 
intermolecular spacing scale or mean free path, then the truncated Euler system 
constitutes a nontrivial model of a molecular liquid, with the equilibrium excitation 
corresponding to normal molecular thermal energy. 
\end{quote}
In the same paper he also suggested that the truncated Euler system could support an
energy cascade just as in the Navier--Stokes system, for ``a statistical ensemble 
whose initial distribution is multivariate-normal, with all energy concentrated in 
wavenumbers the order of $k_0$.''
In 1989 he and S.~Chen  went much further:
\begin{quote}
``the truncated Euler system can imitate
NS fluid: the high-wavenumber degrees of freedom act like a thermal sink
into which the energy of low-wave-number modes excited above equilibrium is
dissipated. In the limit where the sink wavenumbers are very large compared
with the anomalously excited wavenumbers, this dynamical damping acts
precisely like a molecular viscosity.''  
\end{quote}
Actually, in 2005 very-high-resolution spectral simulations of the 3D
Galerkin-truncated Euler equations showed that, when initial
conditions are used that have mostly low-wavenumber modes, the
truncated inviscid system tends at very long times to thermal
equilibrium with energy-equipartition, but it has long-lasting transients 
which are basically the same as for viscous high
Reynolds-number flow, including a K41-type inertial range. In other
words, the high-wavenumber thermalized modes act as an artificial
molecular micro-world.\footnote{The first quote is from Kraichnan, 1975a: 
p.~31 and the second from
Kraichnan \& Chen, 1989: p.~162 {; see also Forster et al., 1977.}
For the simulation of truncated Euler, Cichowlas et al., 2005.} 


\subsection{Two-Dimensional Turbulent Cascades} 
\label{ss:2d}

All this is to set the stage for a major contribution made by Kraichnan 
in 1967, his theory of inverse cascade for two-dimensional
(2D)  turbulence. Equilibrium statistical mechanical reasoning is one 
of the many strands of thought that Kraichnan wove together into a 
compelling argument for the existence of Kolmogorov-type cascades in 
2D incompressible fluids. His remarkable paper in Physics of Fluids invoked
also an analysis of triadic interactions under the assumption of  scale-similarity,
exact results on the instantaneous evolution of Gaussian-distributed  
initial conditions, various plausible statistical and physical arguments, and 
even comparison with Kraichnan's simultaneous work on dynamics of quantum 
Bose condensation. As noted already in the Introduction, Kraichnan's findings have 
considerable interest for understanding large-scale geophysical and planetary flows which, 
due to a combination of small vertical scale, rapid rotation and strong stratification 
may be described by 2D Navier--Stokes or related equations.\footnote{The pioneering paper on 2D turbulent cascades is Kraichnan, 1967b.}

Before discussing Kraichnan's contributions, however, a brief digression on prior 
2D turbulence work  is in order. For many years, following the observation by Taylor 
that there cannot be appreciable energy dissipation in high-Reynolds-number 2D flow,  
two-dimensional turbulence was not considered with much favor.
Vortex stretching was regarded as the essence of turbulence and this effect is absent 
in 2D. Thus, it was often stated---and occasionally still is---that ``there is no 2D turbulence.''  
This situation changed in the late forties, when two-dimensional approximations were
proposed for high-Reynolds-number geophysical flows, which had many of 
the attributes of turbulence (randomness, disorder, etc.)  For example,  J.G.~Charney in 
1948 formulated the quasi-geostrophic model, in which potential vorticity (like vorticity for 
2D Euler) is conserved along every fluid element. 
A little later, J.~von Neumann made a number of interesting remarks 
on 2D turbulence, in particular that it is expected to have far less disorder than in 3D, precisely
because of vorticity conservation; but this material remained unpublished for a long time. 
At the very end of his 1953 turbulence monograph, G.K.~Batchelor proposed to investigate 
the spottiness ``of the energy of high wave-number components'' using  2D turbulence. 
He observed that in a 2D ideal flow the conservation of the integral of (one
half of)  the squared vorticity---Now called ``enstrophy'', a term coined 
by C.E.~Leith, which we shall use liberally---prevents energy from solely flowing to high
wavenumbers: some energy has to be transferred also to smaller wavenumbers (larger scales). 
Batchelor also concurred with Onsager about the tendency of small vortices of the same sign 
to merge into larger vortices. In the same year, R. ~Fj{\o}rtoft showed that, because of the
simultaneous conservation of energy and enstrophy,  it is impossible for 2D dynamics to 
change the amplitudes of only two (Fourier) modes. Within a triad of modes, he showed
that the  change in energy for the member of the triad with intermediate wavenumber is  
the opposite to that of the other two members and that the member with lowest wavenumber
shows the largest energy change of the two extreme members.\footnote{See Taylor, 1917:~pp.~76-77, Charney, 1947; von Neumann, 1949:  \S\S\,2.3--2.4; 
Batchelor, 1953: pp.~186--187; Fj{\o}rtoft, 1953. CROSS REFS. MAY BE NEEDED}

It is thus clear that there must be significant inverse transfer
of energy in 2D. However, even in 3D 
there is some inverse transfer of energy  for the case
of freely decaying high Reynolds number flow, where the peak of
the energy spectrum migrates to smaller wavenumbers. 
What about the K41 energy cascade and the presence of an inertial range over which 
the energy flux is uniform? Lee showed that a \textit{direct} energy cascade is not
possible in 2D, because it would violate enstrophy conservation.\footnote{Lee, 1951.}  
Before 1967 no conjecture is found in the literature positing any type of 2D 
power-law cascade range. 

Then comes Kraichnan. 
The Abstract of his first 1967 paper is worth quoting in full:
\begin{quote}
Two-dimensional turbulence has both kinetic energy and
mean-square vorticity as inviscid constant of motion. Consequently
it admits two formal inertial ranges, $E(k) \sim
\epsilon^{2/3}k^{-5/3}$  and $E(k)\sim \eta ^{2/3} k^{-3}$, where
$\epsilon$ is the rate of cascade of kinetic energy per
uni mass, $\eta$ is the rate of cascade of mean-square vorticity, and the
kinetic energy per unit mass is $\int_0^\infty E(k) dk$. The
$-\frac{5}{3}$ range is found to
entail backward energy cascade, from higher to lower wavenumbers $k$,
together with zero-vorticity flow. The $-3$ range gives an upward
vorticity flow and zero-energy flow. The paradox in these results
is resolved by the irreducibly triangular nature of the elementary
wavenumber interactions. The formal $-3$ range gives a nonlocal
cascade and consequently must be modified by logarithmic
factors. If energy is fed in at a constant rate to a band of
wavenumbers $\sim k_i$  and the Reynolds number is large, it is conjectured
that a quasi-steady-state results with a $-\frac{5}{3}$ range for $k
\ll k_i$ and a $-3$
range for $k\gg k_i$, up to the viscous cutoff. The total kinetic
energy increases steadily with time as the $-\frac{5}{3}$ range pushes to
ever-lower $k$, until scales the size of the entire fluid are strongly
excited. The rate of energy dissipation by viscosity decreases to
zero if kinematic viscosity is decreased to zero with
other parameters unchanged. 
\end{quote}
This is followed by a detailed and very dense presentation of the arguments
supporting such results. Other than the works already cited and known
to Kraichnan, there were no experimental or numerical results which
could guide him. Furthermore, he took up the formidable challenge
of not using closure, although he stated:
\begin{quote}
No use is made of closure approximations. However, the Lagrangian-history 
direct-interaction approximation, which yields Kolmogorov's similarity cascade
in three dimensions, preserves the vorticity constraint in two dimensions
and appears to yield the principal dynamical features inferred in the 
present paper.
\end{quote}
Actually, some of the Fourier-space machinery developed by Kraichnan 
for closure,
such as the use of the energy flux $\Pi(k)$ passing through wavenumber $k$ 
and of the triad contributions 
to energy transfer $T(k,p,q)$, remains meaningful  for homogeneous isotropic turbulence
without recourse to closure. Of course, such quantities must then be
expressed in terms of triple correlations of the velocity field and 
not solely in terms of the energy spectrum. 
Invoking a scaling ansatz for triple correlations, 
$T(ak,ap,aq) = a ^{-(1+3n)/2} T(k,p,q),$  
and also for the energy spectrum,\footnote{
As Kraichnan pointed out, the latter is not used in his argument, thus allowing, in present-day 
language, for anomalous scaling.} $E(k) \propto k^{-n},$
Kraichnan showed
that for $n=5/3$ one obtains 
an inertial range with zero enstrophy flux and non-vanishing energy flux and that for $n=3$ 
one obtains an inertial range with zero energy flux and non-vanishing enstrophy flux. 

These would be the famous dual cascades
of 2D turbulence if Kraichnan could show that the former range has a
negative energy flux, while the latter has a positive enstrophy flux. 
At this point, to predict
the directions of the cascades, he made a very creative use of the equilibrium
statistical mechanics of the Galerkin-truncated Euler equations,
which he now calls ``absolute (statistical) equilibria.''  In 2D, because of the simultaneous
conservation of energy $E$ and enstrophy $\Omega$, it is easy 
to show that when the Euler equations are Galerkin truncated to 
a wavenumber band $[k_{\rm min},\, k_{\rm max}]$ with 
$0 <k_{\rm min}< k_{\rm max}$, the absolute equilibrium 
$e^{-(\alpha E+\beta\Omega)}/Z$ is Gaussian with an energy spectrum
\begin{equation}  
E(k) = \frac{k}{\alpha +\beta k^2} ,
\label{2dabseq}
\end{equation}
where $\alpha$ and $\beta>0$ are constrained by the knowledge (say,
from
the initial conditions) of the total energy $E = \int_ {k_{\rm
    min}}^{k_{\rm max}} E(k) dk$ and of the total enstrophy $\Omega =
\int_ {k_{\rm
    min}}^{k_{\rm max}} k^2 E(k) dk\ge k_{\rm min}^2 E$. If
$\Omega/(k_{\rm min}^2 E)$ is very close to unity, it is seen that the
``inverse temperature'' 
$\alpha$ must be negative and that the spectrum \eqref{2dabseq}
displays a strong peak near  $k_{\rm min}$. 
The situation is in stark contrast to the case of 3D absolute
equilibria: for 3D parity-invariant
flow the absolute equilibrium energy spectrum is proportional to $k^2,$  as we have
seen, and always peaks at the  highest wavenumber. These results suggested to 
Kraichnan that ``a tendency toward equilibrium in an actual physical
flow should involve an upward flow of vorticity and, therefore, by the
conservation laws, a downward flow of energy.'' For good measure, Kraichnan 
gave two independent arguments to justify the same conclusions. Adapting 
earlier considerations of Fj{\o}rtoft, he noted that these cascade directions follow 
if the transfer in each triad is ``a statistically plausible spreading of the excitation 
in wavenumber:  out of the middle wavenumber into the extremes.'' Finally, he 
showed that such  diffusive spreading in wavenumber indeed develops 
instantaneously for an initial Gaussian statistical distribution, applying
expressions of W.H.~Reid and Ogura for the quasi-normal closure in 2D.\footnote{Reid, 1959; Ogura, 1962.}

But are these formal similarity ranges physically realizable? Kraichnan next turned
to this question. Since a $k^{-5/3}$ range gives a divergent energy at small
wavenumbers and the $k^{-3}$ range a logarithmically divergent enstrophy, 
cascade ranges of arbitrary extent require forcing 
at some intermediate wavenumber $k_i$.  For finite ranges, Kraichnan 
noted, there must be some ``leakage'' of energy input $\varepsilon$ to high wavenumbers
and of enstrophy input $\eta$ to low wavenumbers. As either of the ranges 
increases in length it becomes a ``purer'' cascade,  due to the blocking effect  
of conservation of the dual invariant.  The enstrophy cascade will proceed 
up to the cutoff wavenumber $k_d=(\eta/\nu^3)^{1/6}$ set by viscosity, with a 
vanishingly small energy dissipation $\varepsilon_d\sim \varepsilon (k_i/k_d)^2$
for $k_d\gg k_i.$ If there is no minimum wavenumber $k_0$,  Kraichnan concluded 
that an inverse cascade should proceed for ever to smaller and  smaller wavenumbers 
which, on dimensional grounds, scale as a $\varepsilon^{-1/2}t^{-3/2}$, 
where $t$ is  the time elapsed. These cascade ranges are only plausible universal states,
Kraichnan observed,  if the cascade dynamics are scale-local, with the dominant nonlinear 
interactions among triads of wavenumbers,  all of comparable magnitude.
Kraichnan concluded that  the 2D inverse cascade $-5/3$ range is local just as is the 
3D direct cascade $-5/3$ range of Kolmogorov. The 2D direct enstrophy range is at the 
margin of locality, however,  and must have logarithmic  corrections to exact self-similarity. 
Echoing the earlier ideas of von Neumann, Kraichnan argued that the infinite number 
of local vorticity invariants in 2D suggests non-universality of spectral coefficients. 
He then noted
\begin{quote}
A further point is that the nonlocalness of the transfer in the -3 range suggests
in itself that [the] cascade there is not accompanied by degradation of the higher statistics 
in the fashion usually assumed in a three-dimensional Kolmogorov cascade. This is 
consistent with a picture of the transfer process as a clumping-together and coalescence 
of similarly signed vortices with the high-wavenumber excitation confined principally to 
thin and infrequent shear layers attached to the ever-larger eddies thus formed.
\end{quote} 
This is the only point in the paper where Kraichnan speculates on physical-space
mechanisms, clearly influenced by the statistical mechanics argument of Onsager.  

Kraichnan considered finally in his 1967 paper the situation that the fluid is confined to 
a finite domain with a minimum wavenumber $k_0\ll k_i$. He wrote:
\begin{quote}
The conjecture is offered here that after the $-\frac{5}{3}$ range reaches down to 
wavenumbers $\sim k_0$ the downward cascade from $k_i$ continues and the 
energy delivered to the bottom of the range piles up in the mode $k_0.$ As the 
energy in $k_0$ rises sufficiently, modification of the $-\frac{5}{3}$ range toward 
absolute equilibrium is expected, starting at the bottom and working up to 
progressively larger wavenumbers.
\end{quote} 
This conclusion was supported by considering, once again, the 2D
absolute equilibria; for $\Omega/(E k_0^2)$ close to unity,
 Kraichnan showed that they have nearly all of the energy 
carried by the ``gravest'' mode $k_0$. He pointed out an analogy with
quantum 
Bose 
condensation which, apparently, played a key role in stimulating his whole analysis.
Indeed, he wrote in the Introduction:
\begin{quote}
The present study grew out of an investigation of the approach of a weakly 
coupled boson gas to equilibrium below the Bose-Einstein condensation 
temperature. There is a fairly close dynamical analogy in which the number 
density and kinetic-energy density of the bosons play the respective roles 
of kinetic-energy density and squared vorticity.
\end{quote} 
This quantum phenomenon was discussed at length in a prior Kraichnan paper 
in 1967.\footnote{\label{thefootnote} Kraichnan, 1967a}

Two very original concepts enter into turbulence theory with Kraichnan's landmark 
work. The first idea is that a single system may support two, co-existing cascades
with different spectral ranges, in 2D the dual cascades of energy and enstrophy.
The second idea is there may be constant flux spectral ranges that correspond 
to an  \textit{inverse cascade}, from small to large scales. Kraichnan's concept of 
the 2D inverse energy cascade is very far from the Richardson--Kolmogorov vision of 3D
turbulence, in which energy, introduced at large scales either through  the initial conditions 
or by suitable forces or by instabilities, cascades to smaller scales and eventually 
dissipates by viscosity into heat. Two other groups were, however, pursuing ideas 
very closely related to those of Kraichnan, at this same time. These
were G.K.~Batchelor and R.W.~Bray at Cambridge, UK and V.~Zakharov in the Soviet Union.

The 2D enstrophy cascade was proposed independently of Kraichnan, and even 
somewhat earlier, by Batchelor. This result is reported in the 1966 Cambridge PhD 
dissertation of Bray. At the beginning of \S\,1.4 he gave his supervisor, 
Batchelor,  credit for the idea that the enstrophy dissipation rate 
could remain finite in the limit of vanishing viscosity. 
This led Bray to suggest an enstrophy cascade with a 
$k^{-1}$ spectrum and thus a $k^{-3}$ energy spectrum. 
Bray attempted to check this theory by performing a 2D spectral numerical 
simulation, probably the first of its kind. These results were not made public, however, 
until a 1969 paper authored by Batchelor. The analysis of Batchelor and 
Bray is remarkably complementary to Kraichnan's. They considered only 
decaying 2D turbulence, not forced steady states.
Most of their physical discussion was also in real phase, not in spectral space, and 
focused on the analogy between stretching of vorticity-gradients in 2D and Taylor's
vortex-stretching mechanism in 3D. Neither in Bray's thesis nor in Batchelor's 
paper was there any discussion of a separate $-5/3$ range in 2D
with constant flux of energy to large scales (Batchelor was, by 1969,
aware of Kraichnan's earlier paper and cited it in his
work).\footnote{Bray, 1966; Batchelor, 1969. CROSS REFS. MAY BE NEEDED }

The notion of two distinct power-law ranges already appears in the 
1966 PhD thesis of Zakharov, on the weak turbulence of gravity waves. 
One of these ranges was identified as a 
direct cascade of energy to high-wavenumber, but the physical interpretation
of the other was not clearly identified in the thesis. Shortly afterward, 
however, Zakharov hit upon the idea that the second power-law range corresponds 
to an inverse cascade of wave action or ``quasi-particle number.'' 
It thus {appears} that Kraichnan and Zakharov arrived independently
at the idea of an inverse cascade although Kraichnan, it seems, got the idea slightly 
earlier into print.
Both Kraichnan and Zakharov also clearly realized the applicability of these notions 
of dual cascades and inverse cascade to many other systems, including quantum 
dynamics of Bose condensation.\footnote{Zakharov, 1966. B.~Kadomtsev informed Zakharov of Kraichnan's 2D paper 
sometime around 1969 (V. Zakharov, private communication, 2010).}  

%

The subject  of 2D turbulence continued to interest Kraichnan until the early 
1980's, at least, and he wrote several later 
papers which sharpened the predictions and clarified the physics of his 1967 theory. 
In a 1971 J. Fluid Mech. article he applied his TFM closure to both 2D inverse and 
3D direct energy cascades and obtained quantitative results on spectral 
coefficients. 
Kraichnan also studied the 2D enstrophy cascade 
and,  in particular, worked out the logarithmic correction mentioned in 1967.  
A 1976 paper in J.  of Atmosph. Sci. disseminated
these results to the community of 
meteorologists, with special attention to the phenomenological concept of  
``eddy viscosity''.  Kraichnan proposed there a new interpretation of the inverse cascade 
in terms of a negative eddy viscosity, an idea that goes back to V. Starr and he gave a very 
simple heuristic explanation for this effect: 
\begin{quote}
If a small-scale motion has the form of a compact blob of vorticity, or an assembly of 
uncorrelated blobs, a steady straining will eventually draw a typical blob out into an 
elongated shape, with corresponding thinning and increase of typical wavenumber. 
The typical result will be a decrease of the kinetic energy of the small-scale motion 
and a corresponding reinforcement of the straining field.
\end{quote}
This idea has been particularly influential in the geophysical literature, 
where it has often been invoked to explain inverse energy cascade.
\footnote{For Kraichnan's application of TFM closure to 2D turbulence cascades,
see Kraichnan, 1971b, 1976c; on negative viscosity, Starr, 1968;
Kraichnan, 1976c; Kraichnan \& Montgomery, 1980: \S\,4.4;  on vortex-thinning 
mechanism of inverse cascade, Kraichnan, 1976c; Rhines, 1979; Salmon, 1980, 
1998: p.~229. Another notable paper on 2D turbulence is Kraichnan, 1975b.}

What is the empirical status of Kraichnan's dual cascade theory of 2D turbulence? 
A complete review would be out of place here, but we shall briefly discuss its verification 
with an emphasis on the most current work. Only very recently, in fact, has it become 
possible to observe both 2D cascades, inverse energy and direct enstrophy, in a single 
simulation. This has required herculean computations at spatial resolutions 
up to $32,768^2$ grid points.
The simulations confirm Kraichnan's predictions for 
the $k^{-5/3}$ and $k^{-3}$ ranges, with less accuracy for the latter due to finite-range 
effects. Earlier numerical simulations and laboratory experiments which have focussed
on a single range have, however, separately confirmed the predictions of the 1967 paper.  
A number of numerical studies of the enstrophy cascade with
``hyperviscosity'' (powers of the Laplacian replacing the usual
dissipative term) have reported 
observing  the log-correction to the energy spectrum.  
The quasi-steady inverse cascade predicted by Kraichnan 
as a transient before energy from pumping reaches the largest scales has 
also been observed in both experiments and simulations, first by 
{L.~Smith and V.~Yakhot}. The constant flux $k^{-5/3}$ range is cleanly
observed. However, contrary to Kraichnan's speculations in his 1967
paper, the cascade is not associated to ``coalescence'' of vortices
and, indeed, the statistics of the velocity are quite close to
Gaussian and strong, coherent vortices do not appear until the energy
begins to accumulate at the largest scales.  Experiments and
simulations on the statistical steady-state have instead found considerable
evidence for Kraichnan's 1976 ``vortex-thinning'' mechanism of energy
transfer even in the local cascade regime. In
the situation without large-scale damping, there is ``energy
condensation'' at large scales as Kraichnan had supposed, but not
confined to the gravest mode. Recent simulations  show that condensation in a periodic domain appears as a
pair of large, counterrotating vortices with a $k^{-3}$
spectrum. These vortices are close to what is predicted by an
equilibrium, maximum-entropy argument although the system is non-equilibrium,
with continuously growing energy and constant negative energy flux.\footnote{ 
{On numerical simulation of simultaneous cascades, see Boffetta, 2007,
Boffetta \& Musacchio, 2010; on the log-correction in the enstrophy cascade, 
Borue, 1993; Gotoh, 1998; Pasquero \& Falkovich, 2002; on quasi-steady 
cascade, Smith \&Yakhot, 1993; on vortex thinning, Chen et al., 2006; Xiao et al. 2008; 
on condensations, cf. Chertkov et al., 2007; Bouchet \& Simonnet, 2009.}}

Perhaps the most interesting question,
from the general scientific point of view, is the relevance of Kraichnan's ideas to 
planetary atmospheres and oceans. This question is complicated by the limited scale 
ranges that exist in those systems and the greater complexity of the dynamics.
However, several recent observational studies have found evidence for both inverse 
energy and direct enstrophy cascades in the Earth's atmosphere and 
oceans.\footnote{{On direct enstrophy cascade in the Earth's stratosphere, 
Cho \& Lindborg, 2001; on  inverse energy cascade
in the South Pacific,  Scott \& Wang, 2005.}}

\section{Intermittency}
\label{s:intermittency}

Intermittency is a rather general term referring to the spottiness of
small-scale turbulent activity, be it at dissipation-range scales or
at inertial-range scales. In the late forties Batchelor and A.A.~Townsend
observed intermittent behavior of low-order velocity
derivatives;  since such
derivatives come predominantly from the transition region between the
inertial and dissipation ranges, this intermittency cannot be directly
taken as evidence that the self-similarity postulated {
for the K41 inertial-range} 
is breaking down.\footnote{Batchelor \&
  Townsend, 1949.  CROSS REFS. MAY BE NEEDED} 

\subsection{Dissipation-range intermittency}
\label{ss:dri}

Kraichnan was the first to explain
intermittency in the far dissipation range or, equivalently, for
high-order
velocity derivatives. In slightly modernized
form, his argument is as follows: suppose that the flow can be divided
into macroscopic regions each having its energy dissipation
rate $\varepsilon$ and its energy spectrum $E(k)$. In the far
dissipation range $E(k)$ falls off faster than algebraically. From DIA
results or from von Neumann's analyticity conjecture regarding
solutions of the Navier--Stokes equations, Kraichnan
expects a fall-off $\propto \exp(-k/k_d)$, where the dissipation
wavenumber $k_d$ is given, at least approximately, by the K41
expression $(\epsilon/\nu ^3)^{1/4}$. When $k\gg k_d$, even minute
macroscopic fluctuations in $\varepsilon$, which are very likely as
pointed out by {L.D.~Landau}, 
will produce huge macroscopic fluctuations 
in $E(k)$ and thus strong intermittency in physical-space filtered
velocity signals obtained by keeping only those Fourier coefficients which are
in a high-$k$-octave of wavenumbers. This argument can be made more
systematic by using singularities of the analytic continuation of the
velocity field to complex space-time locations.\footnote{Kraichnan, 
1967c, 1974a:~p.~327; von Neumann, 1949; on complex singularities, 
cf. Frisch \& Morf, 1981. }

\subsection{Inertial-range intermittency}
\label{ss:iri}

Much more difficult is the issue of intermittency at inertial-range
scales and the problem of \textit{anomalous scaling}, that is scaling
for which the exponents cannot be obtained by a dimensional argument,
as in K41. In the early sixties, {A.M.~Obukhov and his advisor
Kolmogorov} began to
suspect that K41 must be somewhat modified because spatial
averages $\varepsilon_r$ of the local energy dissipation over balls with a radius $r$
staying within the inertial range appeared to fluctuate more and more
when $r$ is decreased; they proposed a lognormal model of intermittency.\footnote{See
the contribution by Falkovich in this volume. CROSS REFERENCES TO OTHER  PAPERS IN THIS VOLUME NEEDED.}  {E.A.~Novikov and 
R.~Stewart and then A.M.~Yaglom} constructed
ad hoc random multiplicative models to capture such intermittency and
the corresponding scaling exponents. Mandelbrot showed that, in these
models, the dissipation is taking place on a set with non-integer
\textit{fractal} dimension; in general such models are actually
\textit{multifractal}.\footnote{Obukhov, 1962, Kolmogorov, 1962;
Novikov \& Stewart, 1964; Yaglom, 1966; for a review of the Russian
work, cf. Falkovich CROSS REFS. MAY BE NEEDED; Mandelbrot, 1968, 1974; on
multifractality, cf. Parisi \& Frisch, 1985.}

For Kraichnan, who liked proceeding in a systematic way, keeping as much
contact as possible with the true fluid-dynamical equations,
inertial-range intermittency was a very difficult problem. Indeed, it
was
known since 1966 that the full hierarchy of moment or cumulant equations derived
for statistical solutions of the Navier--Stokes equation is compatible
with the scale-invariant K41 theory  in the limit of infinite Reynolds
numbers. But Kraichnan was also aware that K41 is equally compatible
with the Burgers equation, which definitely has no K41 scaling
(because
of the presence of shocks); he also noticed that the presence of the
pressure in the incompressible Navier--Stokes was likely to reduce
the intermittency one would otherwise expect from a simple vortex
stretching argument.  Closure
seemed incapable to say anything about the breaking of the K41 scale
invariance (one major exception to this statement
is discussed in Section~\ref{ss:psi}).\footnote{On K41-compatibility of the Navier--Stokes
equations, cf. Orszag \& Kruskal, 1966; on the differences between
Burgers and Navier--Stokes turbulence, cf. Kraichnan, 1974a, 1991.} 

At first Kraichnan examined critically the toy models developed by the Russian 
school and observed that $\varepsilon_r$ is not a pure inertial-range quantity 
and proposed to study intermittency in terms of more  appropriate
quantities, such as the local fluctuations of the energy flux
associated with a wavenumber $k$ in the inertial range. An estimate of
this flux is $u_r^3/r$, where $r \sim 1/k$ and $u_r$
is, say,  the modulus of the velocity difference between two
points separated by a distance $r$. With this in mind, he wrote:
\begin{quote}
If we increase the intermittency by making the fluid into quiescent
regions with  negligible velocity and active regions, of equal
extent, where  $u_r$ increases by $\sqrt 2$, then the mean kinetic energy in
scales order $r$ is unchanged  but the time constant decreases, and hence 
$\varepsilon$ increases, by $\sqrt 2$. This example suggests, first,
that if Kolmogorov's theory holds in subregions of the fluid, then the
constant $f(0)$ in the inertial-range law can be universal only if
intermittency in the local dissipation $\varepsilon_{r}$, defined as average
dissipation over a domain of size $r$, somewhat tends to a universal
distribution. Second, if intermittency increases as scale size decreases, and
Kolmogorov's basic ideas hold in local regions, then the cascade becomes more 
efficient as $r$ decreases and $E(k)$ must fall off more rapidly than
$k^{-5/3}$ if, according to conservation of energy, the overall cascade rate
is $r$ independent.\footnote{Kraichnan, 1972:~p.~213.}
\end{quote}
A few years later this  remark,
together
with ideas of Mandelbrot, became a key ingredient in the development of 
the $\beta$-model, a
phenomenological
model of intermittency that uses exclusively inertial-range
quantities.\footnote{On using inertial-range quantities, cf. Kraichnan, 1972: p.~213, 1974a; 
Frisch, Sulem \& Nelkin, 1978.}

Kraichnan pursued some of these ideas further himself in an 
influential 1974 paper in J.  Fluid Mech. This paper is 
pure Kraichnan. A wealth of intriguing ideas are tossed out, very original 
model calculations sketched in brief, and clever counterexamples devised 
against conventional ideas. At least two contributions of this paper are now 
well-known. First, Kraichnan proposed a refined similarity hypothesis (RSH)
alternative to that of Kolmogorov, which he based on inertial-range energy
flux rather than volume-averaged energy dissipation. Later numerical and 
experimental studies have confirmed Kraichnan's RSH (and also that of 
Kolmogorov!). In the same paper, Kraichnan gave what is now the standard 
formulation of the ``Landau argument'' on intermittency and non-universality 
of coefficients in scaling laws. His argument is considerably clearer and more
compelling than the brief remarks originally made by Landau in 1942. The crucial 
observation of Kraichnan is that only those K41 predictions which are linear in the 
ensemble-average energy dissipation $\langle \varepsilon\rangle$---such 
as the Kolmogorov 4/5-th law---can be expected to be universally valid 
inertial-range laws. Other K41 predictions which depend upon fractional
powers $\langle \varepsilon\rangle^p$ are not invariant to composition of 
sub-ensembles with distinct global values of mean-dissipation. There is at 
least as much Kraichnan in this argument as there is 
Landau.\footnote{For Landau's argument, Landau \& Lifshitz, 1987;
Kraichnan, 1974a: \S~2; Frisch, 1995: \S~6.4.} 
 
\subsection{Passive scalar intermittency and the ``Kraichnan model''}
\label{ss:psi}

The story of the Kraichnan model and of the birth of the first 
\textit{ab initio} 
derivation of anomalous scaling is rather complex, spanning nearly three
decades. Since it is understood that
in this book the emphasis should be on what happened before 1980, we shall
concentrate on the early developments, that begin in the late sixties. 

The transport of a passive scalar field
(say a temperature field $T({\bm x},\,t)$), advected by a prescribed
incompressible turbulent velocity field ${\bm u}({\bm x},\,t)$ and subject
to molecular diffusion with a diffusivity $\kappa$, is governed by the
following \textit{linear} stochastic equation:
\begin{equation}  
  \left[\partial_t +{\bm u}({\bm x},\,t)\cdot \nabla_x - \kappa \nabla ^2_x\right]T({\bm x},\,t)=0.
\label{passivescalar}
\end{equation}
The qualification ``passive'' is used when there is no or negligible
back-reaction on the turbulent flow of the field being transported.
Examples of passive scalar transported fields are provided by the temperature
of a fluid (when buoyancy is negligible), the humidity of the
atmosphere,
the concentration of chemical or biological species. Passive scalar
transport has thus an important domain of applications and
considerable efforts were made since the forties to gain an
understanding at least as good as for turbulence dynamics.
In particular {Obukhov and, independently, S. Corrsin} 
derived for passive scalars 
the counterpart of the $-5/3$ law, Yaglom derived an analogue
of the Kolmogorov's four-fifths law  and Batchelor derived a $k^{-1}$
passive scalar energy spectrum in a regime of fully developed
turbulence with large Schmidt number (see below).
 It was thus quite natural for Kraichnan to see how well the
closure tools he developed for turbulence in the fifties and the
sixties were able to cope with passive scalar dynamics. 
He applied his LHDIA closure to the passive scalar problem, for example, 
reproducing Obukhov--Corrsin scaling with precise numerical 
coefficients.\footnote{Obukhov, {1949}; Corrsin, 1951; Yaglom, 1949; Batchelor, 1959; 
cf. also the chapters on Batchelor, on Corrsin and on the Russian
School in 
this volume. CROSS REFS. MAY BE NEEDED.
On LHDIA for passive scalars, cf. Kraichnan, 1965a: \S~5--7. }

In 1968 Kraichnan realized that a closed equation can be  
obtained for a scalar field passively advected by a turbulent velocity with a very short 
correlation time, without any further approximation. The DIA closure is exact 
for this special system, reducing to a single equation for the scalar correlation 
function at two space points and simultaneous times. The 
mean Green function reduces to a Dirac delta because of the zero correlation-time
assumption. This 1968 model is now usually called ``the Kraichnan model'' 
[of passive scalar dynamics] and has assumed a paradigmatic status
for turbulence theory, comparable to that of the Ising model in statistical mechanics of  
critical phenomena. Its importance stems from a string of major discoveries 
by Kraichnan and others on the fundamental mechanism of intermittency, some of 
which will be described only briefly because they took place in the nineties.
Kraichnan showed that even  when the velocity  field is not at all intermittent, 
e.g. a Gaussian random field, the passive scalar (henceforth called ``temperature'' for
brevity) can become intermittent and this in several ways.\footnote{Kraichnan, 1968b, 
1974b, 1994.}

A first mechanism, which applies in the far dissipation range, is
basically the same as described in \S\,\ref{ss:dri} and will not
concern us further.

A second mechanism identified by Kraichnan concerns the
so-called Batchelor regime: when the Schmidt number $\nu /\kappa$ is
 large, there is a range of scales for which the velocity field is
strongly affected by viscous dissipation, but the temperature field
does not undergo much diffusion; in this regime the velocity field
can be locally replaced by a uniform random shear.\footnote{Batchelor,
  1959 and Moffatt's contribution in this volume. CROSS REFS. MAY BE NEEDED} Tiny,
well-separated temperature blobs are then stretched and squeezed in a
way which is amenable to asymptotic analysis at large times. Actually
doing this in a systematic way would have required all kinds of heavy-duty
theoretical tools: path integrals, large deviation theory,
fluctuations of Lyapunov exponents, etc.\footnote{See, e.g., the review by Falkovich, Gaw\c{e}dzki \&
Vergassola, 2001.} It is then possible to show that the distributions
of spatial derivatives of the temperature display a lognormal-type
intermittency at zero diffusivity\footnote{This is a non-trivial 
  variant of the obvious result that when $m(t)$ is a scalar Gaussian
 random function the solution of the differential equation $dq(t)/dt = m(t)
q(t)$ with $q(0)=1$  is lognormal.} and a weaker form of intermittency
in the regime with non-zero diffusivity. Actually, all this was
done --- and correctly so --- by Kraichnan in a remarkable paper
published in 1974, just after the paper on  Kolmogorov's
inertial-range theories.\footnote{Kraichnan, 1974b.} 
This paper is a  tour de force, combining
 very original analytical arguments and deep physical intuition
 to reach exact conclusions, without any assistance from the advanced 
 mathematical methods that were later applied to this problem.  
 Kraichnan's analysis was carried out for general space dimension $d$---
 following a suggestion of M. Nelkin---and one intriguing finding was that intermittency 
 of the scalar vanished in the limit $d\rightarrow \infty.$
 Kraichnan's work, which was going to strongly influence
subsequent more formally rigorous analyses,   showed a thorough understanding of 
the mechanism of intermittency in the Batchelor regime. 

The third mechanism identified by Kraichnan was rather close to one of
the Holy Grails of turbulence theory, namely understanding
inertial-range anomalous scaling and predicting the scaling exponents.
In 1994 Kraichnan conjectured that when the velocity  ${\bm u}({\bm
  x},\,t)$ is Gaussian with a power-law spectrum (K41 would be one
instance) and with a very short correlation time (white-in-time), then
for vanishingly small $\kappa$ the structure functions of the
temperature display anomalous scaling. This is a rather amazing
proposal: how can a self-similar velocity field  act on a 
transported temperature field to endow it with anomalous scaling and
thus with lack of self-similarity?  As we shall see, the qualitative aspects of Kraichnan's
conjecture have  been  fully corroborated by later
work.\footnote{Kraichnan, 1994.}

Now we shall have to become slightly more technical to explain how
Kraichnan tackled this problem, starting with his 1968 work. Let us rewrite the temperature
equation
\eqref{passivescalar} in abstract form
\begin{equation}
\partial_t T(t)= M T(t) + \tilde M (t)  T (t),
\label{toy}
\end{equation}
where $M$ is a linear deterministic operator (diffusion) and $\tilde M(t)$
a linear random operator (advection) with vanishing mean and ``very short
correlation time''. More precisely, one performs the substitution
\begin{equation}
\tilde M (t) \longrightarrow \frac{1}{\epsilon} \tilde M 
(\frac{t}{\epsilon^2})
, \qquad \epsilon \rightarrow 0,
\label{limit}
\end{equation}
where $\tilde M(t)$ is statistically stationary. In the limit
$\epsilon \to 0$ the temperature becomes a Markov process (in the time
variable) and it may be shown that the mean temperature satisfies
a \textit{closed} equation, namely\footnote{Hashminskii, 1966, 
Frisch \& Wirth, 1997.}
\begin{eqnarray}
&&\partial_t \langle T \rangle= M \langle T(t) \rangle +
{\cal D} \langle T(t) \rangle, 
\label{mastereq}\\
&& {\cal D} = \int_{0}^{\infty} \langle \tilde M (s) \tilde M (0) \rangle ds.
\end{eqnarray}
Similar closed equations can be derived for $p$-point moments of the
temperature. In 1968 Kraichnan derived the equation for the two-point 
temperature correlation functions by this technique  and found that 
the second-order temperature structure functions displayed scaling. The scaling
exponent $\zeta_2$ can actually be  obtained by simple dimensional
analysis. So far no evidence of anomalous scaling had emerged.

By a {method} similar to that used in 1968 for the
two-point correlations of a passive scalar, Kraichnan derived in 1994 an
equation for the structure
function of order $p$. This
equation
is not closed (contrary to the equation for the $p$-point correlation
function), but Kraichnan proposed a plausible approximate closure ansatz from
which he derived the following scaling exponents $\zeta_p$ for the
$p$th
order structure function:
\begin{equation}  
\zeta_{2p} =\frac{1}{2}\sqrt{4pd\zeta_2-2+(d-\zeta_2)^2} -\frac{1}{2}(d-\zeta_2).  
\label{wrong}
\end{equation}
Since $\zeta_{2p}$ is obviously not equal to $p\zeta_2$, as would be required by
self-similarity, \eqref{wrong}
implies anomalous scaling. One year later it was shown
that there is indeed anomalous scaling, using a \textit{zero modes} method,
borrowed partially from field theory: the equation for the moments of order $2p$ 
has a linear operator $L_{2p}$ acting on the $2p$-point correlation function
and an inhomogeneous right hand side involving  correlation functions
of lower order.  The zero modes correspond to certain functions of $2p$
variables which are killed by $L_{2p}$. 
Actually, determining the zero modes
turned out to be quite difficult. In most instances it could
be done only perturbatively, using as small parameter either
the roughness  exponent $\xi$ of the prescribed
velocity field or the inverse of the dimension of space $d$ 
(as anticipated by Kraichnan's 1974 paper).  The results agreed with 
numerical simulations, but did not agree with \eqref{wrong} except for 
a single value $\xi\doteq 1.$ Kraichnan's prediction \eqref{wrong} must 
cross the numerical curve at one point, trivially, but it is possibly significant 
that Kraichnan's closure ansatz works best in the regime where the cascade 
dynamics is scale-local.\footnote{Kraichnan, 1994.
On zero-mode methods, cf.
Gaw\c{e}dzki \& Kupiainen, 1995; Chertkov et al., 1995; Shraiman \&
Siggia, 1995 and the review by Falkovich, Gaw\c{e}dzki \& Vergassola,
2001. On
simulations, cf. Frisch, Mazzino
\&
Vergassola, 1998; Gat, Procaccia \& Zeitak, 1998; Frisch et al. 1999.
The whole story about anomalous scaling for passive scalars is
recounted in \url{www.oca.eu/etc7/work-on-passive-scalar.pdf}.
}

\section{Miscellany and conclusions}
\label{s:conclusion}


\subsection{Scattering of sound by turbulence}
\label{ss:scattering}

In 1952 M.J.~Lighthill published a landmark paper on the generation of
sound by turbulence.  The next year Kraichnan observed that the
production of noise in this theory depends on a high power of the Mach
number and that the ``\textit{scattering} [of sound by turbulence] is
the most conspicuous acoustical phenomenon associated with very low
Mach number turbulence.'' In his very first published paper, Kraichnan
developed a systematic theory of the interaction of sound with
nearly incompressible turbulence. This paper, together with further
developments was to be the basis of
a nonintrusive ultrasonic  technique for the remote
probing of vorticity. The same year 1953 and independently Lighthill also
published a theory of scattering.\footnote{Lighthill, 1952, 1953;
Kraichnan, 1953; for further developments, see, e.g., Lund \& Rojas,
1989, Ting \& Miksis, 1990; for vorticity probing, see, 
e.g. Baudet, Ciliberto \& Pinton, 1991.}

In his approach to the problem of interaction of sound and turbulence, 
Kraichnan assumed that the turbulence  is incompressible and can be
described by a {divergence-free} velocity, whereas the sound is 
given by a {curl-free} (potential) velocity. As done by Lighthill,
Kraichnan assumed  that density and pressure fluctuations are related
by an adiabatic equation of state with a uniform speed of sound. Starting 
with the full compressible equations he  performed a decomposition 
of the velocity
\begin{equation}  
  {\bm u}  =  {\bm u}^L + {\bm u}^T 
\label{hodge}
\end{equation}
 into a {curl-free} (longitudinal in the spatial Fourier space) and a 
 {divergence-free}
 (transverse in Fourier space) part. (This is known as a  Hodge
 decomposition in mathematics.) He then obtained a wave equation which
 has four terms. One term is linear in ${\bm u}^L$, related to viscous stresses and is mostly
negligible. The three remaining ones are quadratic and of type L-L,
L-T and T-T. The T-T term is Lighthill's (quadrupolar) sound
production 
term. The L-T term gives the scattering of a preexisting sound wave
by the turbulence. Kraichnan then worked out the angular distribution
and frequency distribution of the scattered wave in terms of the
four-dimensional Fourier transform of the shear velocity
field. Explicit expressions for cross sections were obtained for the case
of a scattering from a region of isotropic turbulence. 

Some remarks are in order. Kraichnan worked in relativity 
and quantum field-theory for several
years before engaging in hydrodynamics but this first published
paper is about hydrodynamics;\footnote{His first relativity paper was to
  be published only two years later (Kraichnan, 1955b).} 
  the turbulent flow is here prescribed and
defined as ``characterized by the fact that although the detailed
structure of the system is not known, suitable averages of certain
quantities are known for a representative ensemble of similar
systems.''  The paper is unusually well written for a first paper and
indicates considerable maturity of the young scientist who had
already been active for six years, although he refrained from
publishing.

\subsection{High-Rayleigh number convection}
\label{ss:convection}

Thermal convection is ubiquitous in technology and is amenable
to controlled experiments where a fluid heated from below is placed
between two horizontal plates. Within the so-called Boussinesq
approximation, the  dimensionless parameters are 
the Rayleigh number ${\rm Ra} \equiv g \alpha \delta T h^3 /(\nu
\kappa)$ and the Prandtl number ${\rm Pr} \equiv \nu/\kappa$. Here,
$g$ is the acceleration due to gravity, $\delta T$ the vertical
temperature difference across the fluid of height $h$, and $\alpha$,
$\nu$ and $\kappa$ are the thermal expansion coefficient of the fluid,
its kinematic viscosity and its thermal diffusivity.  Turbulent
thermal convection was and remains a central topic of the Woods
Hole Oceanographic Institute Geophysical Fluid Dynamics summer
program, with which Kraichnan had considerable interaction from the
late fifties. Around the same time he also had much interaction with
E.A.~Spiegel, who had been trained in astrophysical fluid dynamics:
it is usually convective transport which allows the heat generated
in the interiors of stars to escape. In the early days the easiest way
to model astrophysical convection was through the mixing length
theory, which follows ideas of Boussinesq and of Prandtl. 
In 1962 Kraichnan devoted a fairly substantial paper to thermal convection,
which we cannot summarize {in detail} because 
of lack of space. We shall thus concentrate on his most orignal 
contribution, to what is now called ``ultimate
convection'', at  extremely high Rayleigh
numbers.\footnote{Boussinesq, 1870; Prandtl, 1925; Kraichnan, 1962.}

One important question in high-Rayleigh number convection is the
dependence upon Rayleigh and Prandtl numbers of the Nusselt number 
N, the heat flux non-dimensionalized by the conductive heat flux. In the fifties
{C.H.B.~Priestley} found  a dimensional argument 
which suggests that for high Rayleigh
numbers $N \propto {\rm Ra}^{1/3}$. As pointed out by Kraichnan
``[In] Priestley's theory \ldots it  is assumed that at sufficiently
high Rayleigh numbers most of the change in mean temperature across
the layer occurs in thin boundary regions,  at the surfaces, where
molecular
heat conduction and and molecular viscosity are dominant. Elsewhere
\ldots
convective heat transport and eddy viscosity are
dominant.''\footnote{Priestley, 1959 and references therein;
 Kraichnan, 1962:~p.~1374.} 

However, at sufficiently high Rayleigh numbers the thermal boundary
layer may be destroyed and another regime may emerge, which, as shown by
Kraichnan,  has an approximately
square-root
dependence on Ra. This can be partially derived 
by a simple dimensional argument
due to Spiegel, which assumes that the heat flux depends neither
on the viscosity nor on the thermal diffusivity and which gives 
$N \sim ({\rm Ra}{\rm Pr})^{1/2}$.\footnote{It may be shown that this argument
breaks down at large Prandtl numbers.} Kraichnan's derivation makes use of the
phenomenological theory of high-Reynolds number shear flow turbulence near
a solid boundary, which gives a logarithmic correction proportional to 
$(\ln {\rm Ra})^{-3/2}$. Kraichnan also discussed the Prandtl number
dependence of the various regimes. This allowed him to predict how high
the Rayleigh number should be for the square-root regime to dominate:
for a unit Prandtl number this threshold is around ${\rm Ra}= 10^{21}$.
It is generally believed that the threshold is significantly lower and depends
on the Prandtl number and the boundary conditions. Successful attempts 
to observe this law may have been made 
with Helium gas. Artefacts masquerading
as a ${\rm Ra}^{1/2}$ law cannot be ruled out.  In G\"ottingen a two-meter high convection experiment 
using sulfur hexafluoride (SF6) at 20 times atmospheric pressure is under 
construction to try and capture Kraichnan's ultimate convection 
regime.\footnote{Spiegel, 1971. On artefacts, Sreenivasan (private
  communication, 2010). On Helium gas experiments, cf. Chavanne et
 al., 2001; Niemela   et al., 2000. On SF6, cf.  
 {\tt http://www.sciencedaily.com/releases/2009/12/091203101418.htm}}


\subsection{Kraichnan and computers}
\label{ss:computers}

Kraichnan, although basically  a theoretician, was very far from being
allergic to computers. Actually, not only was he a very talented programmer,
but he got occasionally involved in writing system software and even
in modifying hardware.  Some of his closest collaborators, foremost
S.A.~Orszag, prodded by him, got deeply involved in three-dimensional
simulations of Navier--Stokes turbulence. This was --- and still is --- called
``Direct Numerical Simulation'' (DNS) because the original goal was to check
on the validity of various closures by going \textit{directly} to the fluid
dynamical  equations. Convinced that many features of high-Reynolds
number turbulence should be  universal, Kraichnan encouraged the use
of the simplest type of boundary conditions (periodic) which allows
the simple and  efficient use of spectral methods. He also suggested
using Gaussian initial conditions rather than more realistic ones.
Curiously, although the thrust to do DNS started just after Kraichnan's
discovery of the 2D inverse cascade, he strongly recommended focusing
on 3D flow.

Considerable effort --- this time often in collaboration with
J.R.~Herring --- went into the numerical integration of various
closure
equations. Kraichnan proposed using discrete wavenumbers in geometric rather
than arithmetic progression. This allowed reaching very high Reynolds numbers.
Kraichnan himself was actively involved in writing code for these
investigations. His punched cards were shipped from New Hampshire to NASA Goddard
Institute where the machine computations were performed during the 1960's and 1970's.
Herring recalls that ``Bob's programs very rarely contained any bugs.''

\subsection{Conclusions}
\label{ss:conclusions}

Our survey has focussed on three of Kraichnan's contributions to
turbulence theory: (1) spectral closures and realizability, (2)
inverse cascade of energy in 2D turbulence and (3) intermittency of
passive scalars advected by turbulence. These are, arguably, his most
significant achievements which have had the greatest impact on the
field.  Spectral closures of the DIA class still have numerous
interesting applications when the questions under investigation do not
depend crucially on deviations from K41.  Even today an EDQNM
calculation, for example, will often be the first line of assault on a
difficult new turbulence problem. Furthermore, Kraichnan's criterion
of realizability has become part of the standard toolbox of turbulence
closure techniques.  Realizability is necessary both for physical
meaningfulness and, often, for successful numerical solution of the
closure equations. Kraichnan's prediction of inverse cascade has been
well verified by experiments and simulations and has relevance in
explaining dynamical processes in the Earth's atmosphere and
oceans. The concept of an inverse cascade has proved very fruitful in
other systems, too, where similar fluxes of invariants to large-scales
may occur, such as magnetic helicity in 3D MHD turbulence, magnetic
potential in 2D MHD turbulence, and particle number in quantum Bose
systems.  Finally, Kraichnan's model of a passive scalar advected by a
white-in-time Gaussian random velocity has become a paradigm for
turbulence intermittency and anomalous scaling---an ``Ising model'' of
turbulence. The theory of passive scalar intermittency has not yet led
to a similar successful theory of intermittency in Navier--Stokes
turbulence. However, the Kraichnan model has raised the scientific
level of discourse in the field by providing a nontrivial example of a
multifractal field generated by a turbulence dynamics. It is no longer
debatable that anomalous scaling is {\it possible} for
Navier--Stokes.\footnote{On the inverse magnetic helicity,  Frisch et
  al., 1975. On the inverse magnetic potential cascade, cf.  Fyfe \&
  Montgomery, 1976. On Bose condensates, cf. Semikoz
  \& Tkachev, 1997. Regarding intermittency/anomalous scaling, note that
  there have been many incorrect ``proofs'' of their absence in
  Navier--Stokes turbulence, to which the Kraichnan model is a
  counterexample; cf., e.g., Belinicher \& L'vov, 1987.}

A review of Kraichnan's scientific legacy within the length constraint
of this book must be very selective.  For example we have not been
able to discuss the numerous interactions Kraichnan had with many
people in the USA and in other countries, particularly in France,
Israel and Japan. Even focusing our discussion to
his turbulence research prior to 1980, we have been forced to omit
mention of a large number of problems to which Kraichnan made
important contributions in that period. Furthermore, Kraichnan 
invested substantial time to other theoretical approaches, that came 
after the 1980 cutoff for this paper.  We provide 
below just a few references to this additional work on turbulence 
by Kraichnan.\footnote{On the various topics that could not be covered in this 
review, see for pressure fluctuations: Kraichnan, 1956a, 1956b, 1957a;
shear flow turbulence: Kraichnan, 1964a;
magnetic dynamo: Kraichnan \& Nagarajan, 1967; Kraichnan, 1976a,
1976c, 1979b;
Vlasov plasma turbulence: Kraichnan \& Orszag, 1967c;
predictability and error growth: Kraichnan, 1970b; Kraichnan \& Leith, 1972; 
Burgers: Kraichnan, 1968a, 1999; Kraichnan \& Gotoh, 1993;   
quantum turbulence: Kraichnan, 1967a;
path-integrals: Kraichnan, 1958a:~\S\,4.3;  Lewis \& Kraichnan, 1962;
self-consistent Langevin models: Kraichnan, 1970c; 
variational approaches: Kraichnan, 1958a:~\S\,4.3; Kraichnan, 1979a;
Wiener chaos expansions: Kraichnan, 1979a;
Pad\'e approximants: Kraichnan, 1968c, 1970a;
decimation:  Kraichnan, 1985, 1988;
mapping closure: Kraichnan et al., 1989, Kraichnan, 1991; 
Kraichnan \& Kimura, 1993; Kraichnan \& Gotoh, 1993; 
critique of Tsallis statistics for turbulence: Gotoh \& Kraichnan,
2004.}  It may be that later
generations will find that our survey has missed some of Kraichnan's
most significant accomplishments. The richness of his \oe{u}vre can only
be appreciated by poring over his densely written research articles,
bristling with original ideas and novel methods, for oneself. The
reader who does so will be generously rewarded for his effort.
  
It is amusing to wonder what might be Einstein's assessment (from the
welkins) of his former assistant. He would probably have to conclude
that Kraichnan had a lot of {\it Sitzfleisch}.\footnote{The Germans
  have aptly called Sitzfleisch the ability to spend endless hours at
  a desk doing grueling work. Sitzfleisch is considered by
  mathematicians to be a better gauge of success than any of the
  attractive definitions of talent with which psychologists regale us
  from time to time (Gian-Carlo Rota).} Kraichnan's papers, numbering
more than a hundred and spanning five decades, many of them formidably
technical, bear ample witness to their author's iron
determination and staying
power. Turbulence is a dauntingly difficult subject
  where any significant advance is won by a hard-fought battle; and yet
  Kraichnan has left his record of victories throughout the field.
Several  international  conferences  held in his honor  are testimony
to the lasting impact of Robert H. Kraichnan.\footnote{Los Alamos (May
  1998) for Kraichnan's 70th birthday and Santa Fe (May 2009)
and Beijing (September 2009) after he left us in 2008.},

{\bf Acknowledgements.} Many have helped us with their remarks and their own recollections. We are particularly
indebted to B.~Castaing,  {C.~Connaughton}, G.~Falkovich, H.~Frisch, T.~Gotoh, 
J.R.~Herring,  C.E.~Leith, H.K.~Moffatt, S.~Nazarenko, S.A.~Orszag, I.~Procaccia, 
H.~Rose, E.A.~Spiegel, K.~Sreenivasan, B.~Villone and V.~Zakharov. 
{GE's work was partially supported by NSF Grant Nos. AST-0428325 \& CDI-0941530
and UF's by COST Action MP0806 and by ANR ``OTARIE'' BLAN07-2\_183172.}

\end{document}